\documentclass[sigconf,natbib=true]{acmart}
\AtBeginDocument{%
  }

\usepackage{algorithm}
\usepackage{algorithmic}
\usepackage{graphicx}
\usepackage{subcaption}
\usepackage{multirow}
\usepackage[normalem]{ulem}
\useunder{\uline}{\ul}{}

\begin{document}

\title{Towards End-to-End Alignment of User Satisfaction via Questionnaire in Video Recommendation}

\author{Na Li}
\authornote{These authors contributed equally to this work.}
% \affiliation{
%   \institution{Harbin Institute of Technology}
%   \city{Harbin}
%   \country{China}
% }
\affiliation{
  \institution{Kuaishou Technology}
  \city{Beijing}
  \country{China}
}
\email{24s003043@stu.hit.edu.cn}

\author{Jiaqi Yu}
\authornotemark[1]
\affiliation{
  \institution{Kuaishou Technology}
  \city{Beijing}
  \country{China}
}
\email{yujiaqi03@kuaishou.com}

\author{Minzhi Xie}
\authornotemark[1]
\affiliation{
  \institution{Kuaishou Technology}
  \city{Beijing}
  \country{China}
}
\email{xieminzhi@kuaishou.com}

\author{Tiantian He}
\authornotemark[1]
\affiliation{
  \institution{Kuaishou Technology}
  \city{Beijing}
  \country{China}
}
\email{hetiantian05@kuaishou.com}

\author{Xiaoxiao Xu}
\authornotemark[1]
\affiliation{
  \institution{Kuaishou Technology}
  \city{Beijing}
  \country{China}
}
\email{xuxiaoxiao05@kuaishou.com}

\author{Zixiu Wang}
\affiliation{
  \institution{Kuaishou Technology}
  \city{Beijing}
  \country{China}
}
\email{wangzixiu@kuaishou.com}

\author{Lantao Hu}
\affiliation{
  \institution{Kuaishou Technology}
  \city{Beijing}
  \country{China}
}
\email{hulantao@kuaishou.com}

\author{Yongqi Liu}
\affiliation{
  \institution{Kuaishou Technology}
  \city{Beijing}
  \country{China}
}
\email{liuyongqi@kuaishou.com}

\author{Han Li}
\affiliation{
  \institution{Kuaishou Technology}
  \city{Beijing}
  \country{China}
}
\email{lihan08@kuaishou.com}

\author{Kaiqiao Zhan}
\affiliation{
  \institution{Kuaishou Technology}
  \city{Beijing}
  \country{China}
}
\email{zhankaiqiao@kuaishou.com}

\author{Kun Gai}
\affiliation{
  \institution{Unaffiliated}
  \city{Beijing}
  \country{China}
}
\email{gai.kun@qq.com}

% \affiliation{%

% \affiliation{%


\begin{abstract}
Short-video recommender systems typically optimize ranking models using dense user behavioral signals, such as clicks and watch time. However, these signals are only indirect proxies of user satisfaction and often suffer from noise and bias. 
Recently, explicit satisfaction feedback collected through questionnaires has emerged as a high-quality direct alignment supervision, but is extremely sparse and easily overwhelmed by abundant behavioral data, making it difficult to incorporate into online recommendation models.
To address these challenges, we propose a novel framework which is towards \textbf{E}nd-to-End \textbf{A}lignment of user \textbf{S}atisfaction via \textbf{Q}uestionaire, named \textbf{EASQ}, to enable real-time alignment of ranking models with true user satisfaction. 
Specifically, we first construct an independent parameter pathway for sparse questionnaire signals by combining a multi-task architecture and a lightweight LoRA module. 
The multi-task design separates sparse satisfaction supervision from dense behavioral signals, preventing the former from being overwhelmed. 
The LoRA module pre-inject these preferences in a parameter-isolated manner, ensuring stability in the backbone while optimizing user satisfaction. 
Furthermore, we employ a DPO-based optimization objective tailored for online learning, which aligns the main model outputs with sparse satisfaction signals in real time. 
This design enables end-to-end online learning, allowing the model to continuously adapt to new questionnaire feedback while maintaining the stability and effectiveness of the backbone.
Extensive offline experiments and large-scale online A/B tests demonstrate that EASQ consistently improves user satisfaction metrics across multiple scenarios. 
EASQ has been successfully deployed in a production short-video recommendation system, delivering significant and stable business gains.

\end{abstract}


% \ccsdesc[500]{Computing methodologies~Supervised learning}



\maketitle

\section{Introduction}
Short-video sharing platforms have witnessed explosive growth, making recommender systems a critical component that directly impacts user experience and drives the platform’s commercial success \cite{zhao2023video1, xu2025video2, wei2024video3}. 
In particular, ranking models \cite{bai2022mmoe, tang2020ple, pi2020sim,zhou2019dien, wang2025home} play a central role by selecting the most suitable content from a set of candidates, aiming to maximize user satisfaction. 
To achieve this, models rely on carefully designed optimization objectives, which directly shape the final ranking.

Most existing ranking models are trained on relatively dense and easily collected user behavioral signals, such as clicks and watch time \cite{zhou2018din,zhou2019dien,pi2020sim,hidasi2015rnn,zhang2024saqrec}.
However, these signals are only proxy indicators of user satisfaction. Because users express their satisfaction in different ways, these proxies can be noisy or biased, resulting in suboptimal model performance.
Although some prior studies (e.g., EMER \cite{he2025emer}) attempt to incorporate relative satisfaction to refine the supervision signal, the majority of methods still lack direct supervision indicating whether users are truly satisfied.

To better understand and improve user satisfaction, recommender systems have begun to introduce questionnaires \cite{zhang2024saqrec,bakhshi2025retentive}. Questionnaires require users to provide explicit, direct satisfaction signals, which can then provide a high-quality optimization objective for ranking model. 
However, such signals are extremely sparse in real world, and they are easily overwhelmed by a large amount of behavioral signals under conventional training. 
Based on data collected from our real-world recommendation environment, questionnaires are issued for only about 0.5\% of total video views. Moreover, with a click-through rate of less than 2\%, only a tiny fraction of questionnaires yield valid responses, making satisfaction signals extremely sparse and difficult to leverage for effective training.

Existing research has made various attempts to mitigate the problem of sparse satisfaction signals. 
SAQRec \cite{zhang2024saqrec} designs a satisfaction model to predict user satisfaction levels. 
Within the reinforcement learning framework, reward shaping \cite{christakopoulou2022imputation,li2025q2sp} is used to address the problem of sparse rewards.
However, these methods rely on offline-trained auxiliary models or pre-defined reward functions, which are then plugged into online or reinforcement learning frameworks.
This two-stage paradigm limits the ability of the system to continuously adapt to new questionnaire feedback in an online manner.

Inspired by recent advances in preference optimization and alignment in large language models \cite{rafailov2023dpo,ouyang2022rlhf,chen2024dporecommend}, we seek to achieve end-to-end alignment of ranking model outputs with true user satisfaction in an online learning setting.
To overcome these issues, we develop a framework named \textbf{EASQ}, short for \textbf{E}nd-to-End \textbf{A}lignment of user \textbf{S}atisfaction via \textbf{Q}uestionnaire, aiming to achieve real-time alignment between ranking models and true user satisfaction.
% To this end, we construct a dedicated parameter pathway for sparse user satisfaction signals and design a DPO loss specifically tailored for online alignment, naming the framework \textbf{EASQ}.
Specifically, we first establish a decoupled multi-task architecture with independent expert networks for main task and satisfaction alignment task. At the backbone’s lower layers, we introduce a lightweight LoRA module to pre-inject semantic preferences during the embedding stage while preventing sparse questionnaire signals from being overwhelmed by abundant behavioral data.
Through the coordination between the lower-level LoRA and the upper-level expert networks, an independent parameter pathway is created for sparse questionnaire signals.
Furthermore, considering the synchronous updates inherent in online learning, we treat the questionnaire expert network as an online surrogate of a reference model and employ DPO to further align the main model with the sparse questionnaire signals.
EASQ enables end-to-end integration of sparse satisfaction signals in an online learning setting, resulting in ranking models that are directly aligned with real user satisfaction.

Our contributions are summarized as follows:
\begin{itemize}
    \item We propose \textbf{EASQ}, the first framework to directly integrate sparse questionnaire-based satisfaction signals into online learning. The framework constructs an independent parameter pathway, enabling real-time alignment of ranking models with true user satisfaction.
    \item We design a multi-task modeling mechanism with a lightweight LoRA module at the backbone’s lower layers and dedicated expert networks, along with a DPO-based optimization objective, enabling effective alignment of sparse satisfaction signals while preserving the backbone’s stability and capturing human value judgments.
    \item We conduct extensive system experiments and online A/B testing on large-scale industrial data to verify the effectiveness of the proposed framework, and it is stably integrated into a real online recommendation system, demonstrating significant business benefits.
\end{itemize}

\section{Related Work}
\subsection{User Satisfaction}
User satisfaction, which serves as the core metric for evaluating recommendation quality, has attracted widespread attention in recent years.

Some studies \cite{xu2024us1,cai2023us2,chen2023us3} attempt to characterize user satisfaction using behavioral signals that are closer to users’ real experiences. 
For example, BatchRL-MTF \cite{zhang2022BatchRL-MTF} adopts dwell time and the positive-interaction rate as proxy signals for long-term satisfaction. 
PDA \cite{zhang2021pda} argues that explicit “like” feedback is more closely aligned with satisfaction objectives than clicks. 
EMER \cite{he2025emer} addresses the difficulty of quantifying absolute satisfaction by defining relative satisfaction based on posterior behavioral feedback and optimizing AUC. 
While these approaches can partially improve user experience, they rely on indirect signals and therefore struggle to accurately capture users’ true subjective feelings.

Another line of work focuses on directly collecting explicit satisfaction data to obtain higher-quality supervision signals. 
Imputation Network \cite{christakopoulou2022imputation} requires users to provide explicit ratings for consumed content. 
Several studies, including SAQRec \cite{zhang2024saqrec} and prior work from Google \cite{christakopoulou2020google}, collect users’ subjective satisfaction through questionnaires and demonstrate that explicit survey signals better align with users’ true preferences than behavioral feedback. 
In addition, some studies \cite{he2024xinli} draw on insights from psychology and adopt structured measurement scales to assess users’ subjective experiences for more reliable and interpretable evaluation.

Overall, while behavioral signals offer advantages in terms of scale and ease of collection, directly collected questionnaire-based feedback provides a more accurate representation of user satisfaction. 
Consequently, how to effectively leverage sparse yet high-value questionnaire-based satisfaction data has attracted increasing attention in recent research.

\subsection{User Satisfaction Modeling}
In user satisfaction modeling, existing studies primarily focus on improving the quality of supervision under the extreme sparsity of explicit satisfaction signals.

Existing studies predict user satisfaction by explicitly modeling users’ subjective experiences. 
Some works \cite{chen2023us3,liu2024us4} rely on auxiliary behavioral signals to approximate satisfaction, while more recent approaches attempt to infer users’ latent satisfaction states directly from observed interactions.
Models such as Imputation Network \cite{christakopoulou2022imputation} and SAQRec \cite{zhang2024saqrec} aim to reconstruct missing satisfaction labels by learning mappings from behavioral feedback to latent satisfaction representations, thereby alleviating the sparsity of explicit satisfaction signals.
Research based on causal inference also offers new perspectives. 
Some work \cite{schnabel2018shortlong} corrects the inherent response bias in questionnaire signals through inverse propensity weighting (IPW), while methods such as CounterCLR \cite{wang2023counterclr} employ counterfactual inference to estimate user feedback in unobserved or under-exposed data.
Reinforcement learning–based approaches typically treat satisfaction as sparse reward.
To mitigate optimization challenges caused by this, prior work introduces reward shaping or intermediate surrogate objectives, such as Q2SP \cite{li2025q2sp}, which decomposes satisfaction into more frequent sub-rewards.
With the rapid development of large language models (LLMs), recent studies (e.g., OAIF \cite{guo2024oaif}) leverage LLMs as preference annotators, obtaining preference signals that are more closely aligned with subjective experiences through pairwise comparison–based judgments.

However, most existing approaches still rely on behavior-based proxies, strong causal assumptions, or offline supervision, making it difficult to stably align sparse satisfaction signals with dominant behavioral feedback, particularly in online or continual learning settings, especially under dynamic online learning scenarios.

% \verb|\vspace| not allowed

% \begin{figure*}[t]
%     \centering
%     \includegraphics[width=\textwidth]{pdf/main.pdf}
%     \caption{From Conventional Ranking Optimization to Online Learning with User Satisfaction.
% (a) Conventional online learning framework.
% (b) Proposed EASQ framework that aligns recommendations with sparse satisfaction feedback.}
%     \label{fig:framework}
% \end{figure*}

\begin{figure*}
    \begin{subfigure}{0.49\linewidth}
        \centering
        \includegraphics[width=\linewidth]{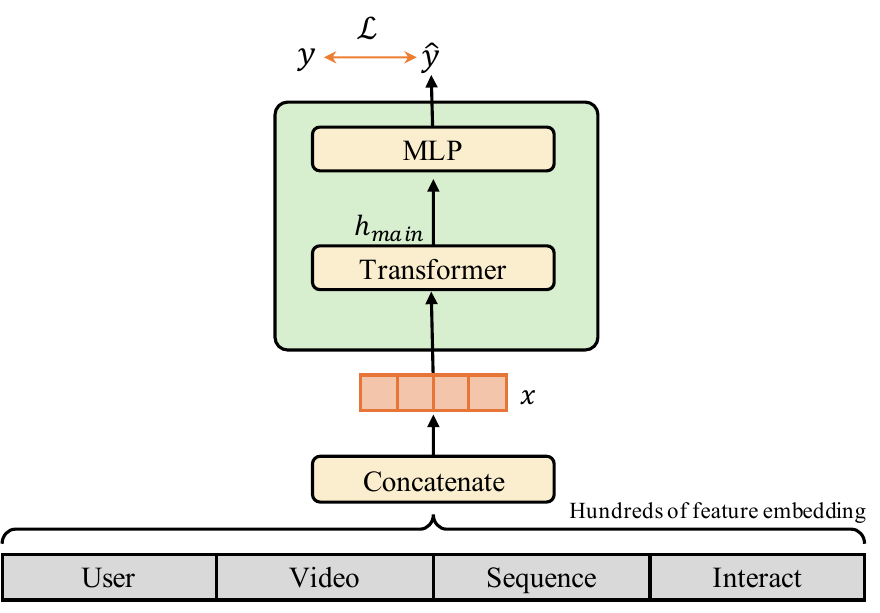}
        \caption{Conventional Framework}
    \end{subfigure}
    \hfill
    \begin{subfigure}{0.49\linewidth}
        \centering
        \includegraphics[width=\linewidth]{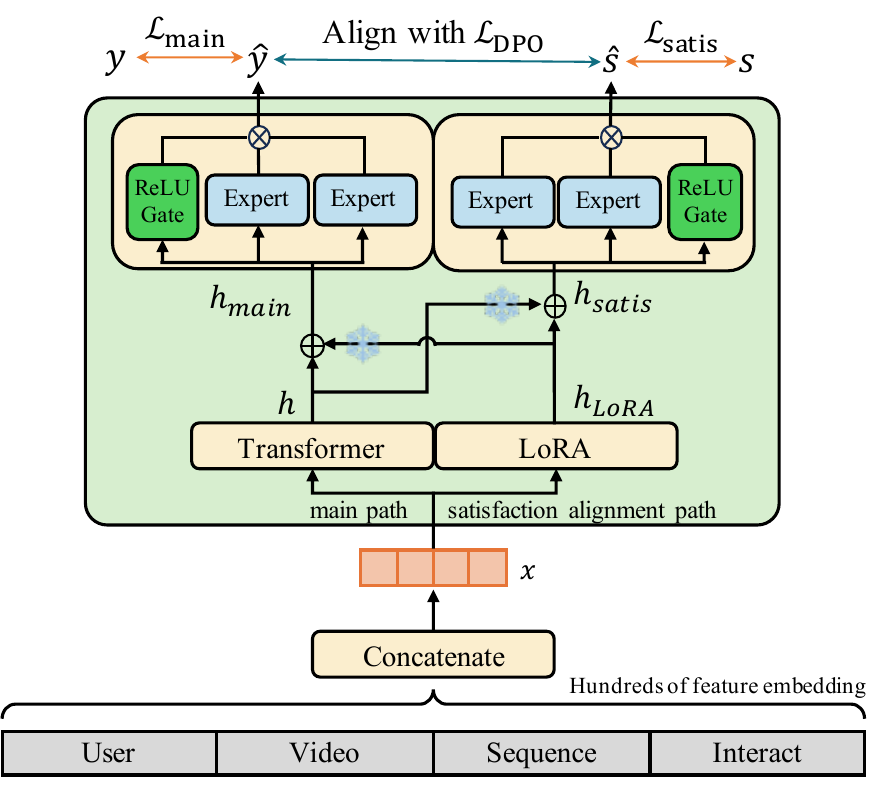}
        \caption{Our Framework EASQ}
    \end{subfigure}
    \caption{From Conventional Ranking Optimization to Online Learning with User Satisfaction.
(a) Conventional online learning framework.
(b) Proposed EASQ framework that aligns recommendations with sparse satisfaction feedback.}
    \label{fig:framework}
\end{figure*}

\section{Preliminary}
\subsection{Problem Statement}
Let $\mathcal{U}=\{u_1,u_2,...\}$ denote the set of users and $\mathcal{X}=\{x_1,x_2,...\}$ denote the set of items in large-scale platforms.
For a given user $u \in \mathcal{U}$, we define a behavioral set $\mathcal{B}_u$, which contains various implicit interaction behaviors of user $u$ on items in $\mathcal{X}$, such as click, watch, like, or skip.
In addition, we define a questionnaire-based feedback set $\mathcal{S}_u$, which consists of explicit satisfaction signals collected from user $u$ through questionnaires on a subset of items in $\mathcal{X}$.
Compared with behavioral signals, $\mathcal{S}_u$ is sparse but of high quality, as it directly reflects the user’s true satisfaction.

Given both $\mathcal{B}_u$ and $\mathcal{S}_u$, our goal is to learn a ranking function 
$f(u, x)$ that can effectively leverage dense behavioral signals while being explicitly aligned with sparse questionnaire feedback, so as to produce a carefully ranked list of items that maximizes the true satisfaction of user $u$.

\subsection{Direct Preference Optimization}
Recent work in natural language processing (NLP) has investigated leveraging human-labeled pairwise comparisons as a reward signal to align language models (LMs) with human preferences, such as RLHF \cite{ouyang2022rlhf} and DPO \cite{rafailov2023dpo,chen2024dporecommend}.
RLHF introduces a preference network to capture the distribution of human choices. 
By modeling preference distribution, a reward function is subsequently learned through maximum likelihood estimation over the comparison data.
Building on previous work \cite{chen2024sdpo}, we adapt it to the recommendation domain. 
The training objective is defined as follows:
\begin{equation}
    \mathcal{L}=-\log p\left(x^+ \succ x^- \mid u\right),
    \label{eq:rlhf}
\end{equation}
where $x^+$ is a preferred item and $x^-$ is a less preferred item. $u$ can represent user profile.
The model is guided by learning preference pairs in a pairwise ranking manner.

Let $\pi_{\theta}(x \mid u)$ denote the probability of recommender producing item $x$ given user $u$.
In the final reinforcement learning stage, the goal is to maximize the expected reward of policy while not deviate too far from the reference model, leading to the following objective for optimal policy:
\begin{equation}
    \max _{\pi_{\theta}} \mathbb{E}\left[r\left(u, x\right)\right]-\beta \mathbb{D}_{\mathrm{KL}}\left[\pi_{\theta}\left(x \mid u\right) \| \pi_{\mathrm{ref}}\left(x \mid u\right)\right],
    \label{eq:rlhf2}
\end{equation}
where $r(\cdot)$ represents reward function and $\beta$ is a temperature coefficient. $\pi_{\mathrm{ref}}$ represents a pre-trained reference model obtained from existing behavioral data, serving as a baseline for preference alignment and stabilizing optimization dynamics. 

DPO \cite{rafailov2023dpo,chen2024dporecommend} theoretically proves the optimal policy in a closed form to Eq.~\ref{eq:rlhf2} is
\begin{equation}
    \pi^{*}\left(x \mid u\right)=\frac{1}{Z\left(u\right)} \pi_{\mathrm{ref}}\left(x \mid u\right) \exp \left(\frac{1}{\beta} r\left(u, x\right)\right),
\end{equation}
which is equivalent to
\begin{equation}
    r\left(u, x\right)=\beta \log \frac{\pi\left(x \mid u\right)}{\pi_{\mathrm{ref}}\left(x \mid u\right)}+\beta \log Z\left(u \right),
    \label{eq:dpo}
\end{equation}
where $Z\left(u\right)=\sum_{x} \pi_{\mathrm{ref}}\left(x \mid u\right) \exp \left(\frac{1}{\beta} r\left(u, x\right)\right)$ is the partition function for normalization in practice.

We can define $p\left(x^+ \succ x^- \mid u\right)$ as $\sigma\left(r\left(u, x^+\right)-r\left(u, x^-\right)\right)$ in Eq.~\ref{eq:rlhf} according to the BT model used in RLHF and substitute term in Eq.~\ref{eq:rlhf} with Eq.~\ref{eq:dpo}. 
Given $u$, the last two phases of RLHF pipeline can be equivalently transformed into optimizing DPO loss below:
\begin{equation}
    \mathcal{L}_{\mathrm{DPO}}=-\log \sigma\left(\beta \left[ \log \frac{\pi_{\theta}\left(x^+\right)}{\pi_{\mathrm{ref}}\left(x^+ \right)}-\log \frac{\pi_{\theta}\left(x^- \right)}{\pi_{\mathrm{ref}}\left(x^- \right)}\right]\right),
\end{equation}
wherein $\sigma(\cdot)$ is the sigmoid function.

DPO is able to directly extract the optimal policy from pairwise preference data.
Nevertheless, DPO relies on a fixed reference model, which is not suitable for end-to-end online learning. We design EASQ to address this issue.  

\section{Method}
In this section, we present the key components and design principles of \textbf{EASQ}.
We first discuss how questionnaire-derived feedback is collected and processed, which provide sparse but informative feedback for guiding model updates.
We then detail the mechanisms for integrating these sparse signals into the backbone to achieve real-time alignment with user satisfaction.

\subsection{Overall Framework}
Unlike conventional model shown in Figure~\ref{fig:framework}(a), the core idea of \textbf{EASQ} is to construct a relatively parameter-independent and stable alignment pathway for sparse satisfaction signals, preventing them from being overwhelmed by behavioral signals during online updates in large-scale systems. 
Figure~\ref{fig:framework}(b) illustrates the overall framework of EASQ.
We first detail the design and collection process of questionnaire-based satisfaction signals.
Then, we introduce a lightweight LoRA module at the lower layers of the model, which injects sparse satisfaction information at an early stage. 
At the upper layers, we further prepare decoupled multi-task learning for sparse questionnaire with main task and satisfaction alignment task. 
We also employ an online-tailored DPO objective to further align the backbone outputs with sparse satisfaction signals, ensuring real-time adaptation during online learning.
By combining these modules, EASQ enables the coordinated progression of behavioral learning and satisfaction alignment.

\begin{figure}
    \centering
    \includegraphics[width=\linewidth]{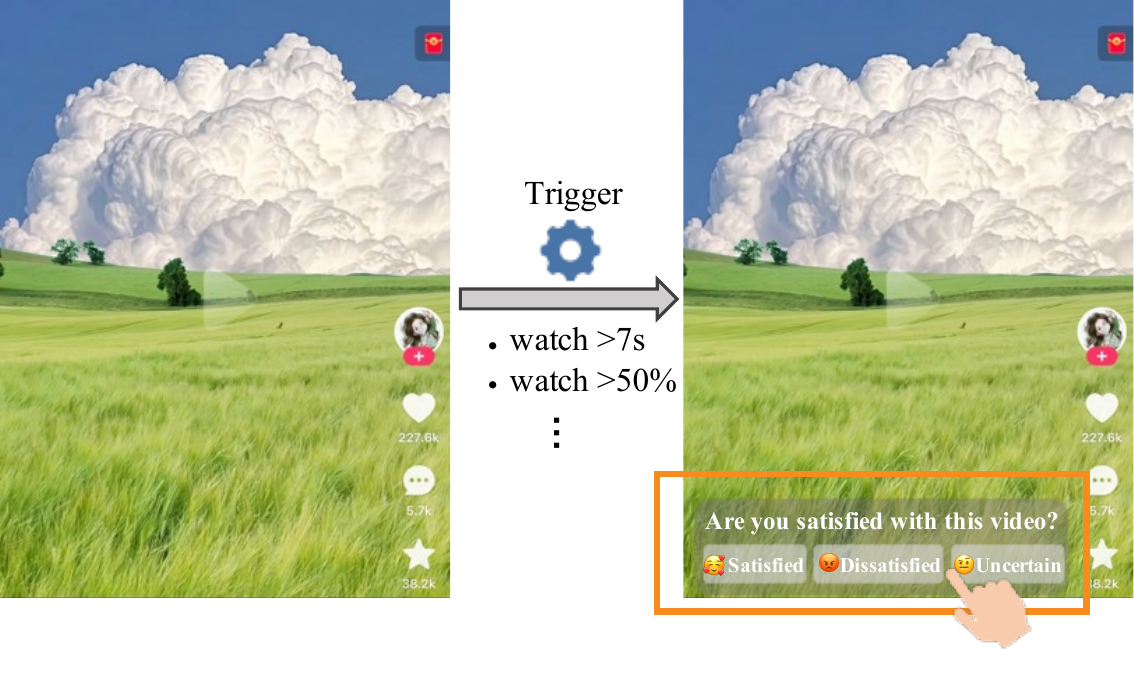}
    \caption{Illustration of questionnaire triggering and implementation.}
    \label{fig:questionnaire}
\end{figure}

\subsection{Satisfaction Signals}
Users’ posterior feedback on items, such as likes and dwell time, can reflect certain dimensions of satisfaction \cite{zhang2022BatchRL-MTF,zhang2021pda}. 
However, even when multiple feedback sources are available, it remains difficult to define the absolute level of user satisfaction based solely on behavioral signals. 
We seek feedback that truly represents users’ satisfaction. From the standpoint of user experience, explicit user evaluations constitute the most direct and faithful expression of satisfaction. 
Accordingly, we incorporate questionnaire-based explicit satisfaction feedback as the key supervision signal for the alignment pathway.

\subsubsection{Questionnaire Theoretical Basis.}
We design the questionnaire to capture users’ expressions of satisfaction based on their experiences with the video content. 
This design is grounded in the theory of experienced utility \cite{kahneman1997experiencedutility}, which posits that directly eliciting individuals’ subjective experiences is an effective way to assess their true feelings.
A key distinction between satisfaction signals and other forms of feedback that focus on interest profiling or value alignment lies in their more direct evaluative semantics and stronger temporal immediacy. 
For example, interest-based questionnaires tend to capture users’ relatively long-term interest structures, while value-alignment questionnaires reflect users’ underlying values and attitudinal preferences. In contrast, satisfaction questionnaire signals explicitly reflect users’ immediate experience quality with the current video, making them more closely aligned with the true user satisfaction that our model aims to predict.

\subsubsection{Questionnaire Design.}
We adopt a direct satisfaction questionnaire that explicitly asks users, \textit{“Are you satisfied with this video?”}. 
Directly eliciting satisfaction enables a more accurate capture of users’ holistic evaluations of content quality. 
The questionnaire provides three response options: \textit{Satisfied}, \textit{Dissatisfied}, and \textit{Uncertain}, where the \textit{Uncertain} option reflects users’ lack of confidence in their judgment, thereby reducing noise introduced by forced choices.

\subsubsection{Questionnaire Implementation.}
To reduce user interaction cost, we embed a lightweight questionnaire at the bottom of the short-video playback interface, as shown in Figure ~\ref{fig:questionnaire}. 
Users can provide feedback with a single click during viewing, without redirection or interrupting the playback flow. 
This design effectively captures users’ immediate experiences with the current recommended content \cite{csikszentmihalyi1987validity}.
To mitigate potential biases, we carefully design the trigger mechanism. 
% Specifically, questionnaires are randomly deployed on popular and relatively recent content , which helps reduce interference caused by content quality. 
The questionnaire is triggered only after users have watched at least 7 seconds of the video or reached 50\% of the playback progress, ensuring that the feedback is grounded in a meaningful viewing experience.
guaranteeing that feedback reflects a meaningful and representative viewing experience.

\begin{figure}
    \begin{subfigure}{0.49\linewidth}
        \centering
        \includegraphics[width=\linewidth]{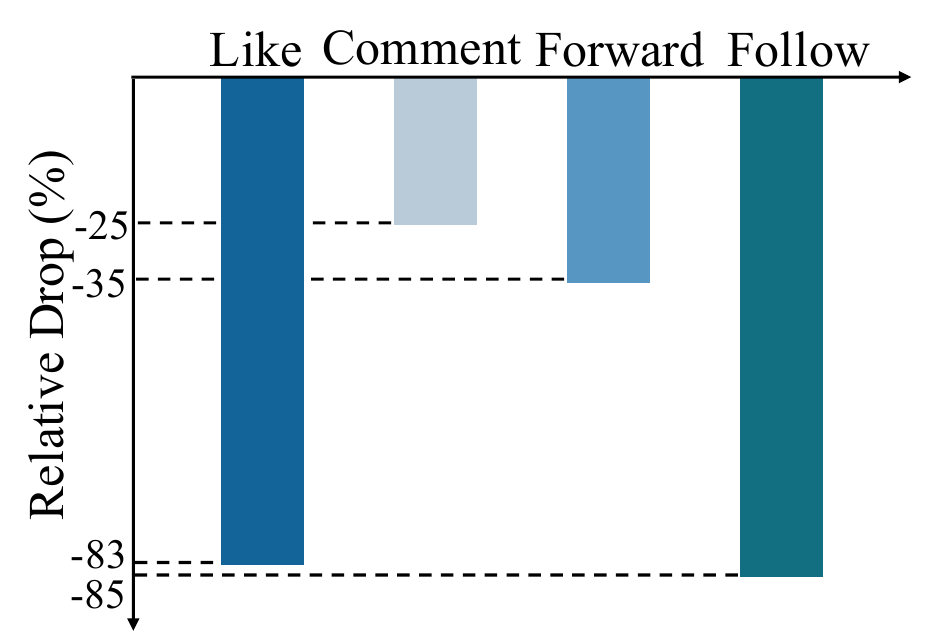}
        \caption{Behavior Drop on Unsatisfied}
        \label{fig:drop}
    \end{subfigure}
    \hfill
    \begin{subfigure}{0.49\linewidth}
        \centering
        \includegraphics[width=\linewidth]{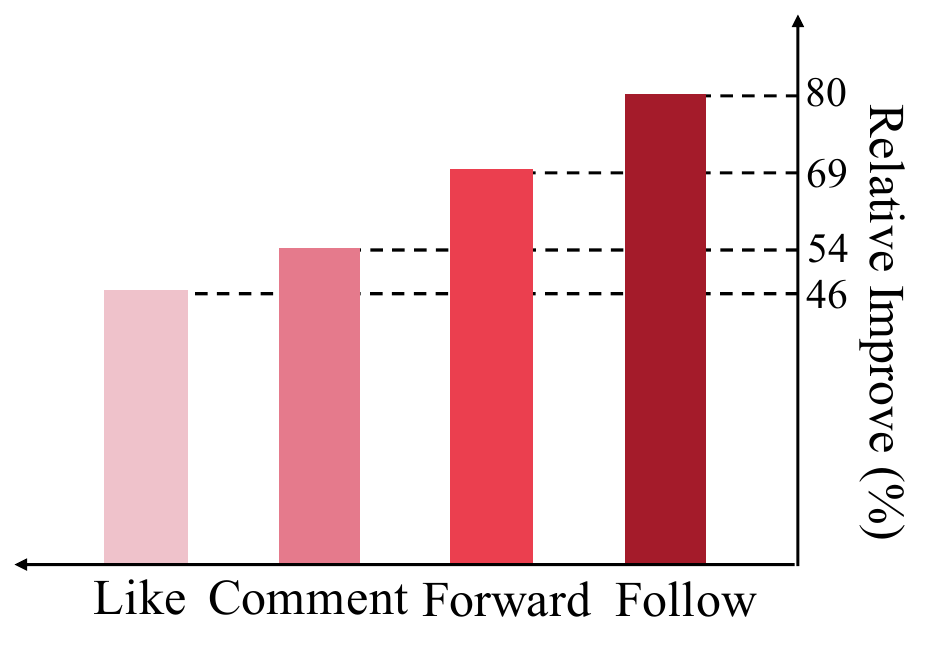}
        \caption{Behavior Improve on Satisfied}
        \label{fig:improve}
    \end{subfigure}
    \caption{Correlation Between Questionnaire Feedback and Posterior User Behavioral Signals.}
    \label{fig:validity}
\end{figure}

\subsubsection{Validity Verification.}
Convergent validity \cite{campbell1959convergent1,cohen2013convergent2} is used to assess whether theoretically related measurement methods exhibit strong positive correlations while still maintaining clear conceptual boundaries. 
To evaluate convergent validity, we analyze the consistency between questionnaire feedback and posterior user behaviors.
As shown in Figure \ref{fig:validity}(a), users who select \textit{Dissatisfied} in the questionnaire exhibit significantly lower posterior behavioral signals on the corresponding videos compared to their average behavior across all videos, whereas \textit{Satisfied} responses are associated with significantly higher-than-average posterior behaviors as shown in Figure \ref{fig:validity}(b). 
The results indicate a strong correlation between questionnaire-based satisfaction signals and posterior behavioral feedback, demonstrating good convergent validity. 
This supports the use of questionnaire signals as reliable supervision for satisfaction alignment modeling.

It is worth noting that the questionnaire-based satisfaction signals are extremely sparse. 
On one hand, issuing questionnaires inevitably introduces interruption to user experience \cite{bakhshi2025retentive}, resulting in a low exposure rate of questionnaires—approximately 0.5\% of total video views. 
On the other hand, the questionnaire click-through rate is less than 2\%, which leads to an extremely limited number of collected responses. Such extreme sparsity of satisfaction signals poses significant challenges for effective model training.

% \begin{figure}
%     \centering
%     \includegraphics[width=\linewidth]{pic/validity.png}
%     \caption{validity}
%     \label{fig:validity}
% \end{figure}

\subsection{Parameter Fine-tuning}
High-value explicit satisfaction signals are extremely sparse in practice.
During the overall training process, they often fail to receive sufficient gradient weight to drive large-scale parameter updates due to this. 
Furthermore, in an online learning setting where parameters are updated simultaneously, it is necessary to ensure that sparse satisfaction signals can be integrated without disrupting the main model.

To address those challenges, we employ LoRA \cite{hu2022lora,kong2024lora2} to construct an independent parameter pathway, enabling separate modeling of sparse satisfaction signals while keeping the main model parameters relatively stable. 
LoRA introduces a small number of trainable low-rank matrices, allowing the model to rapidly incorporate sparse feedback without incurring significant computational overhead.

Let the output of the main model, which can be instantiated with any backbone architecture, be denoted as
\begin{equation}
    h=W_0x,
    \label{eq:h}
\end{equation}
where $x$ represents the input features of both the user and candidate item sides.
For the linear projection matrices $W_0\in \mathbb{R}^{d\times k}$ that require adaptation, LoRA introduces a low-rank update represented as $\Delta W=BA$, where $A\in \mathbb{R}^{r\times k}$ and $B\in \mathbb{R}^{d \times r}$ are trainable low-rank matrices with rank $r\ll \text{min}(d,k)$, substantially reducing the parameter overhead.
Based on this formulation, the incremental representation produced by LoRA for the main model output is given by
\begin{equation}
    h_{LoRA}=\Delta Wx =BAx,
    \label{eq:h_lora}
\end{equation}
effectively injecting sparse satisfaction information into backbone while preserving the stability and robustness of the main model.
% After being inserted into the lower layers of the model, the output of LoRA is added to the main model representation, forming a fused embedding that incorporates user satisfaction information:
% \begin{equation}
%     h=h_{main}+h_{LoRA}.
% \end{equation}
% This design allows sparse satisfaction signals to exert a controlled influence during embedding and lower-layer modeling, without directly modifying the backbone parameters of the main model.

The LoRA module injects semantic preferences from sparse questionnaire signals into the lower layers of the backbone through parameter-efficient fine-tuning. 
Without this early-stage adaptation, the sparse signals risk being overwhelmed by abundant behavioral data, limiting their influence on upper layers.
With LoRA, the subsequent Multi-Task module can focus on modeling satisfaction signals more effectively.
Ablation results show that LoRA consistently improves performance, confirming its necessity for leveraging sparse questionnaire feedback.
% However, since LoRA is inherently a lightweight structure, its expressive capacity is limited. 
% On one hand, LoRA shares parts of the computation graph with the backbone model. As a result, its updates can still be influenced by gradients dominated by dense behavioral signals, which may partially overwhelm the satisfaction signals.
% On the other hand, applying excessively strong LoRA updates may weaken the backbone model’s ability to predict core behavioral objectives. This limitation is further validated by our ablation studies: when using LoRA alone, improvements in satisfaction-related metrics are very limited. 
% To address these limitations, we introduce the MoE-SFT module on top of the LoRA structure, constructing a more independent alignment pathway that allows sparse satisfaction supervision to exert a stable influence.

\subsection{Decoupled Multi-Task Learning for Sparse Questionnaire}
On top of the LoRA-enhanced backbone, we design a MoE-Align module dedicated to sparse satisfaction signals. 
To better capture sparse questionnaire signals, we adopt a multi-task architecture above the backbone, providing a dedicated parameter pathway that enables separate and effective modeling of these signals.
By constructing parallel pathways of main task and satisfaction alignment task at the upper layers of the backbone, we structurally decouple sparse satisfaction signals from the main task learning objectives. 
This design enables sparse signals to be modeled in a relatively independent and stable subspace. In the following, we describe the two pathways in detail.

\subsubsection{Main Task.}
For the main task, the output of LoRA is added to the main model representation, forming a fused embedding that incorporates user satisfaction information:
\begin{equation}
    h_{main}=h+\text{stop\_grad}(h_{LoRA}),
    \label{eq:h_main}
\end{equation}
where $\text{stop\_grad} (\cdot)$ blocks backward gradients from flowing into the LoRA module.
In this way, the main task can benefit from the information provided by LoRA, but does not participate in updating its parameters.
This decoupled design, which resembles the separation between policy and reward modeling in LLM alignment paradigms, enables the backbone to utilize satisfaction-aware features while maintaining stable training during online learning.

The fused embedding is then fed into the MoE component, which implements the main task through multiple expert networks.
Each expert is a feed-forward network (FFN) \cite{vaswani2017transformer} specialized in capturing certain latent preferences.
Specifically, we include $K_1$ FFNs as task experts, and the prediction output for the main task is obtained by linearly combining the outputs of all experts according to a gated routing mechanism:
\begin{equation}
    \hat{y}=\sum_{k}^{K_1} R(h)_{k}\cdot \text{FFN}_k(h),
\end{equation}
where $R(\cdot)$ denotes the routing function.
The main task branch employs dynamic gated routing, which adaptively generates expert weights based on the current input. 
Most prior works \cite{shazeer2017topkrouter1,zoph2022topkrouter2} adopt a Top-K Softmax router, normalizing the gating scores via softmax and selecting the top-K experts by hard truncation. While effective, this introduces a non-differentiable step.

\begin{figure}[h]
    \centering
    \includegraphics[width=\linewidth]{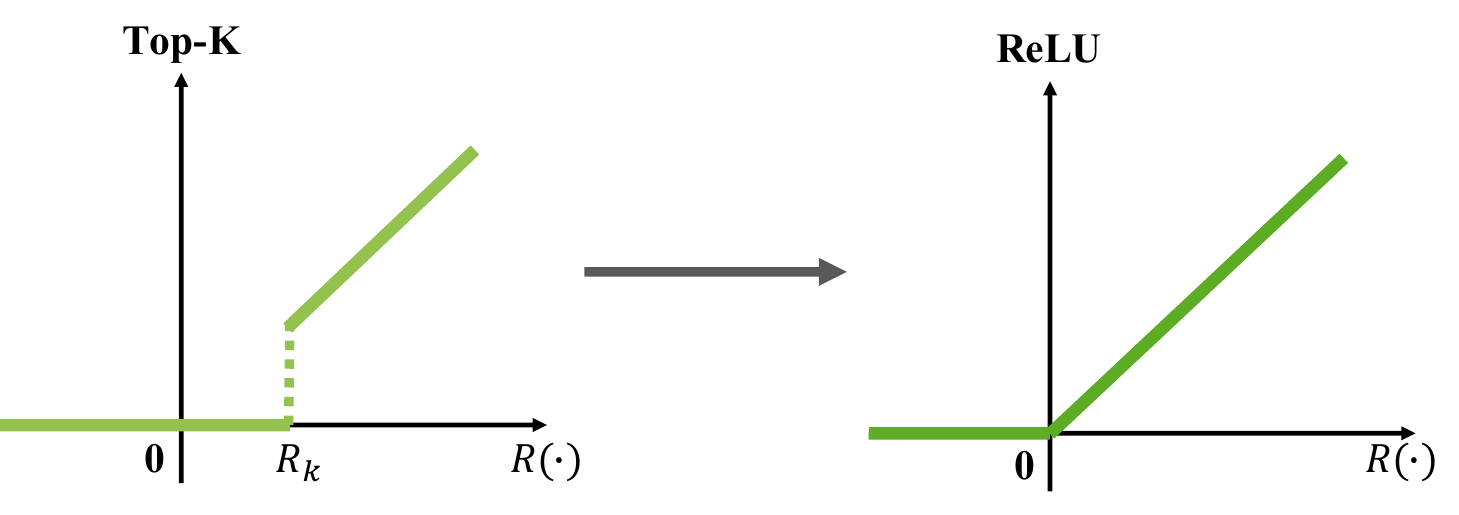}
    \caption{Top-K Softmax vs. ReLU Expert Routing.}
    \label{fig:gating}
\end{figure}

We adopt a ReLU Router as a replacement \cite{wang2024relu}.  
The ReLU Router uses non-negative activations, allowing weights to be zero, thereby enabling fully differentiable learning to activate or suppress experts. 
As shown in Figure ~\ref{fig:gating}, the Top-K router forces each token to be routed to a fixed number of experts.  
In contrast, the ReLU Router allows tokens to be routed to a variable number of experts, significantly improving the flexibility of expert utilization. 
Specifically, the routing function is defined as:
\begin{equation}
    R(h)=\text{ReLU}(W_{gate}h),
\end{equation}
where $W_{gate}$ is the trainable gating matrix.
Finally, the outputs of all activated experts are aggregated to produce the main task prediction.

% Following the loss formulation used in EMER \cite{he2025emer}, we adopt BPR, a widely established objective for pairwise learning from implicit feedback， which encourages items with stronger preferences to receive higher predicted scores than items with weaker preferences:
Following EMER \cite{he2025emer}, we adopt BPR \cite{rendle2012bpr}, a standard objective for pairwise learning from implicit feedback, which enforces higher scores for preferred items over less preferred ones:
\begin{equation}
\begin{aligned}
\mathcal{L}_{\text {main}} & =-\sum_{(i, j) \in \mathcal{B}^+} \log \left(P\left(x_{i} \triangleright x_{j}\right)\right) \\
& =-\sum_{(i, j) \in \mathcal{B}^+} \log \left(\operatorname{sigmoid}\left(\hat{y}_{i}-\hat{y}_{j}\right)\right),
\label{eq:l_main}
\end{aligned}
\end{equation}
where $\mathcal{B}^+=\{(i,j)| y_i>y_j\}$ indicates that item $x_i$ is preferred over item $x_j$ according to observed behavioral signals $y$.
% \begin{equation}
%     P_{i,j}=\text{sigmoid}(\hat{y}_i-\hat{y}_j),
% \end{equation}
% \begin{equation}
%     \mathcal{L}_{\mathrm{pxtr}}=-\sum_{(i,j)}y_{pxtr_{ij}}\log(P_{i,j})+(1-y_{pxtr_{i,j}})\log(1-P_{i,j}),
% \end{equation}
% where $pxtr$ corresponds to different task-specific targets.
% Following EMER \cite{he2025emer}, we assign self-evolving weights $w_{pxtr}$ to different targets:
% \begin{equation}
%     \mathcal{L}_{\mathrm{main}}=-\frac{1}{N}\sum_{pxtr}^{pxtrs}w_{pxtr}\cdot \mathcal{L}_{\mathrm{pxtr}},
%     \label{eq:l_main}
% \end{equation}
% where $N$ is the total number of $pxtr$.

\subsubsection{Satisfaction Alignment Task.}
For satisfaction alignment task, to ensure that it only perceives the satisfaction information injected via LoRA without destabilizing the main model parameters, we construct a partially frozen input:
\begin{equation}
    h_{satis}=\text{stop\_grad}(h) + h_{LoRA},
    \label{eq:h_satis}
\end{equation}
where $stop\_grad(\cdot)$ is used to block backward gradients, preventing the sparse satisfaction signals from updating the main model parameters.
This design allows the satisfaction alignment pathway to independently absorb sparse signals while avoiding conflicts with gradients from the main task.

We employ an alignment expert network $MoE_{align}$, which has the same architectural configuration as the main task pathway but a different number of experts.
Similar to the main task, we construct $K_2$ experts using feed-forward networks (FFNs) \cite{vaswani2017transformer} to produce the alignment output $\hat{s}$:
\begin{equation}
    \hat{s}=\sum_{k}^{K_2} R(h_{satis})_{k}\cdot \text{FFN}_k(h_{satis}).
\end{equation}

Consistent with main task, satisfaction alignment task also uses the ReLU activation function:
\begin{equation}
    R(h_{satis})=\text{ReLU}(W_{gate}  h_{satis}).
\end{equation}

For the satisfaction alignment prediction loss, the sparse satisfaction signals are used in the ranking stage, so the model focuses more on the relative ordering between samples.
Accordingly, we adopt a pairwise logistic loss to encourage all positively ordered pairs $\hat{s}_{i}$ to be ranked higher than $\hat{s}_{j}$:
\begin{equation}
    \mathcal{L}_{\mathrm{satis}} = -\sum_{(i,j)\in D^+}\log(\text{sigmoid}(\hat{s}_{i}-\hat{s}_j)),
    \label{eq:l_satis}
\end{equation}
where the set of positive pairs is defined as $D^+=\{(i,j)|s_i>s_j\}$ and $s_i$ denotes the label derived from the questionnaire.

\subsection{Satisfaction Alignment}

The goal of satisfaction alignment is not only to fit sparse questionnaire preferences, but also to transfer such preferences into the main ranking space in a stable manner. 
Directly optimizing the main model with sparse pairwise feedback may lead to unstable updates, especially in an online learning setting where parameters are updated continuously.

To achieve stable preference alignment, we adopt the principle of Direct Preference Optimization (DPO) \cite{rafailov2023dpo,chen2024dporecommend}, which aligns a target model with human preferences by comparing it against a reference model. 
The reference model serves as an anchor that preserves reasonable behavior while allowing the target model to move toward preferred outputs. 
Traditional DPO relies on a frozen pretrained reference model and a static preference dataset, which makes it difficult to incorporate fresh feedback and respond to rapid changes in user preferences.
In online learning, user interests and satisfaction signals evolve continuously, and models are expected to update in near real time. 
Under this setting, a fixed reference model quickly becomes outdated, and the preference data used by DPO no longer reflects the current distribution.

Therefore, we designed a DPO alignment mechanism specifically adapted for online learning.
Instead of relying on a frozen pretrained reference model, we treat the questionnaire expert network as an online surrogate of the reference model, dynamically reflecting users’ current satisfaction patterns.
During online training, the main model is optimized not only with conventional behavioral signals but also with satisfaction pairs constructed from questionnaire. 
In this way, DPO is no longer tied to static offline data, but becomes an alignment process driven by fresh user feedback.

Given a preferred item $x^+$ and a less preferred item $x^-$ derived from questionnaire feedback $s$, the DPO objective is defined as:
% \begin{equation}
%     \mathcal{L}_{\mathrm{DPO}} = -\log \sigma\left( \left[ \log \frac{\pi_\theta(x^+)}{\pi_\theta(x^-)} 
% - \log \frac{\pi_{\text{ref}}(x^+)}{\pi_{\text{ref}}(x^-)} \right] \right),
%     \label{eq:l_dpo}
% \end{equation}
\begin{equation}
    \mathcal{L}_{\mathrm{DPO}}=-\log \sigma\left(\beta \left[ \log \frac{\pi_{\theta}\left(x^+\right)}{\pi_{\mathrm{ref}}\left(x^+ \right)}-\log \frac{\pi_{\theta}\left(x^- \right)}{\pi_{\mathrm{ref}}\left(x^- \right)}\right]\right),
    \label{eq:l_dpo}
\end{equation}
% where $\pi_\theta$ denotes the main model, $\pi_{\text{ref}}$ denotes the reference model approximated by the frozen satisfaction alignment network.
where $\beta$ is fixed to 0.1 following common practice. $\pi_\theta(\cdot)=\hat{y}$ and $\pi_{\text{ref}}(\cdot)=\text{stop\_grad}(\hat{s})$. $\pi_\theta$ denotes the main model, and $\pi_{\text{ref}}$ denotes the reference model approximated by the satisfaction alignment network.
During back-propagation of $\mathcal{L}_{\text{DPO}}$, gradients are only applied to $\pi_\theta$, while the parameters of $\pi_{\text{ref}}$ are detached from this loss.
The satisfaction alignment network is instead updated separately using its own objective $\mathcal{L}_{\text{satis}}$.

\begin{algorithm}[t]
\caption{The Optimization Process of EASQ}
\renewcommand{\algorithmicrequire}{\textbf{Input:}}
\renewcommand{\algorithmicensure}{\textbf{Output:}}
\begin{algorithmic}[1]
    \REQUIRE Dataset with satisfaction feedback $\mathcal{D}$, base model $f(\cdot)$ with parameters $W$, augmented parameters $\theta$, satisfaction weight $\lambda$, alignment weight $\beta$;
    \ENSURE Updated parameters $W$ and $\theta$.
    \STATE Initialize main model for behavioral signals
    \STATE Initialize satisfaction model for sparse satisfaction signal
    \FOR{each online recommendation batch}
        \STATE Compute backbone output $h$ based on Eq.~(\ref{eq:h})
        \STATE Compute LoRA output $h_{LoRA}$ based on Eq.~(\ref{eq:h_lora})
        \STATE Construct input $h_{main}=h+\text{stop\_grad}(h_{LoRA})$
        \STATE Compute main loss $\mathcal{L}_{\text{main}}$ based on Eq.~(\ref{eq:l_main})
        \STATE Construct input $h_{satis}=\text{stop\_grad}(h) + h_{LoRA}$
        \STATE Compute satisfaction loss $\mathcal{L}_{\text{satis}}$ based on Eq.~(\ref{eq:l_satis})
        \STATE Compute DPO loss $\mathcal{L}_{\text{DPO}}$ based on Eq.~(\ref{eq:l_dpo})
        \STATE Compute total loss $\mathcal{L}_{\text{total}}=\mathcal{L}_{\text{main}}+\lambda_1\mathcal{L}_{\text{satis}}+\lambda_2\mathcal{L}_{\text{DPO}}$
        \STATE Update parameters $W$ and $\theta$ by minimizing $\mathcal{L}$
    \ENDFOR
\end{algorithmic}
\end{algorithm}

EASQ avoids the limitations of a frozen reference model, and transforms sparse, noisy questionnaire feedback into reliable signals of true user satisfaction.
As a result, satisfaction alignment is continuously transferred from the questionnaire task to the main ranking task in a stable and online manner.

The overall training objective is a weighted combination of the two parts:
\begin{equation}
    \mathcal{L}_{\mathrm{total}} = \mathcal{L}_{\mathrm{main}}+\lambda_1\mathcal{L}_{\mathrm{satis}}+\lambda_2\mathcal{L}_{\mathrm{DPO}},
    \label{eq:l_total}
\end{equation}
where $\lambda_1$ controls the contribution of the satisfaction modeling task, which is trained as an independent auxiliary objective, and $\lambda_2$ controls the strength of the alignment objective based on DPO.

It is important to emphasize that the model output used for online inference is the main task prediction $\hat{y}$.
The satisfaction alignment pathway serves only as an auxiliary alignment module during training, continuously absorbing sparse signals and refining the underlying LoRA representations. 
This enables the backbone model to gradually align with true user satisfaction without compromising performance on the core task.

\begin{table*}
\caption{Performance comparisons between EASQ and the baselines across two scenarios.The best and second-best performance methods are denoted in bold and underlined fonts, respectively.}
\label{tab:main results}
\begin{tabular}{@{}cc|cccccc@{}}
\toprule
Scenario                    & Model              & NDCG@5           & NDCG@10          & HR@1             & HR@5             & HR@10            & MRR              \\ \midrule
\multirow{4}{*}{Scenario\#1}& EMER$_\mathbf{S}$  & 0.3347           & 0.3529           & 0.4375           & 0.7813           & 0.9063           & 0.6021           \\
                            & Imputation Network & 0.3479           & 0.3653           & 0.4459           & 0.8046           & \textbf{0.9082*} & 0.6022           \\
                            & SAQRec             & {\ul 0.3577}     & {\ul 0.3704}     & {\ul 0.4688}     & {\ul 0.8125}     & 0.9063           & {\ul 0.6041}     \\
                            & EASQ        & \textbf{0.3729*} & \textbf{0.3887*} & \textbf{0.4773*} & \textbf{0.8286*} & {\ul 0.9081}     & \textbf{0.6294*} \\ \midrule
\multirow{4}{*}{Scenario\#2}& EMER$_\mathbf{S}$  & 0.3474           & 0.3558           & 0.4699           & 0.8551           & 0.9384           & 0.6279           \\
                            & Imputation Network & 0.3553           & 0.3641           & 0.4487           & 0.8550           & {\ul 0.9453}     & 0.6166           \\
                            & SAQRec             & {\ul 0.3563}     & {\ul 0.3687}     & {\ul 0.4807}     & {\ul 0.8578}     & 0.9434           & {\ul 0.6499}     \\
                            & EASQ        & \textbf{0.3784*} & \textbf{0.3812*} & \textbf{0.5111*} & \textbf{0.8668*} & \textbf{0.9503*} & \textbf{0.6683*} \\ \bottomrule
\end{tabular}
\end{table*}

\section{Experiment}

\subsection{Experimental Setup}

\subsubsection{Dataset.}
Our model is deployed in the ranking stage of a real-world industrial short-video platform.
The dataset is collected from practical business scenarios and consists of historical logs from two different usage scenarios.
All data are generated from real user interactions during daily platform usage, faithfully reflecting authentic recommendation processes and feedback behaviors in large-scale industrial environments.

\subsubsection{Evaluation Metrics.}
We evaluate our method using a set of widely used ranking metrics, including Hit Ratio (HR), Normalized Discounted Cumulative Gain (NDCG), and Mean Reciprocal Rank (MRR).
The reported results include HR@\{1, 5, 10\}, NDCG@\{5, 10, 20\}, and MRR.
The evaluation is conducted on positive–negative item pairs derived from questionnaire-based feedback.
All metrics are computed based on the ranked lists of candidate items, and the final performance is obtained by averaging over all users.

\subsubsection{Baselines.}
% (1) \textbf{EMER}: an end-to-end ensemble ranking model using transformers to capture comparative relationships among items.
% (1) \textbf{EMER}$_\mathbf{S}$: a variant of EMER which incorporates questionnaire feedback as supervision for ranking.
% (2) \textbf{Imputation Network}: a model with an additional head to predict user satisfaction from questionnaire feedback.
% (3) \textbf{SAQRec}: a user satisfaction method leveraging imputed questionnaire feedback to handle sparsity and selection bias.
We evaluate the model’s performance against a set of representative benchmark methods, which cover both behavior-driven ranking models and approaches that explicitly incorporate questionnaire-based satisfaction signals.
Specifically, we compare our method with the following models:
\begin{itemize}
    \item \textbf{EMER} \cite{he2025emer}: an end-to-end ensemble ranking model using transformers to capture comparative relationships.
    \item \textbf{EMER}$_\mathbf{S}$: a variant of EMER which incorporates questionnaire feedback as supervision for ranking.
    \item \textbf{Imputation Network} \cite{christakopoulou2022imputation}: a model with an additional head to predict user satisfaction from questionnaire feedback.
    \item \textbf{SAQRec} \cite{zhang2024saqrec}: a user satisfaction method leveraging imputed questionnaire feedback to handle sparsity and selection bias.
\end{itemize}
To ensure a fair comparison under the setting with questionnaire supervision, we reimplemented the above baselines on top of EMER \cite{he2025emer} framework and adapted them to support datasets containing questionnaire signals.

\subsubsection{Implementation Details.}
For different types of features, we set their embedding dimensions to 64, 32, and 8, respectively.
The model uses a multi-head attention module as the backbone, with Softplus as the activation function and L2 normalization to ensure numerical stability.  
The training is performed using the Adam optimizer \cite{kingma2014adam} with an initial learning rate of $1 \times 10^{-3}$.

\begin{table}[]
\caption{Online A/B performance comparison between EASQ and the Baseline Method. $\downarrow$ denotes that a lower value of the corresponding metric is better.}
\label{tab:online results}
% \resizebox{\linewidth}{!}{
\begin{tabular}{lcc}
\hline
                               & Scenario\#1                             & Scenario\#2                             \\
\multirow{-2}{*}{Metrics}      & \multicolumn{1}{l}{EASQ vs Online} & \multicolumn{1}{l}{EASQ vs Online} \\ \hline
LT7                            & +0.042\%                                & +0.043\%                                \\
AppStayTime                    & +0.401\%                                & +0.344\%                                \\
WatchTime                      & +0.563\%                                & +0.456\%                                \\
VideoView                      & +0.634\%                                & +0.840\%                                \\
Like                           & +0.497\%                                & +0.243\%                                \\
Follow                         & +0.253\%                                & +0.301\%                                \\
Comment                        & +0.179\%                                & +0.978\%                                \\
Forward                        & +2.177\%                                & +2.052\%                                \\
Q-rate                         & +1.601\%                                & +1.692\%                                \\
Q-Satisfied                    & +0.929\%                                & +0.531\%                                \\
Q-Dissatisfied $\downarrow$    & -0.652\%                                & -1.129\%                                \\ \hline
\end{tabular}
\end{table}

\subsection{Overall Performance}
As shown in Table ~\ref{tab:main results}, we compare EASQ with aforementioned baselines on our dataset.
From the  results, we can observe that:

(1) Firstly, EASQ consistently achieves the best performance across all evaluation metrics.
Compared with strong behavioral baselines and satisfaction-aware methods, EASQ obtains the highest scores on most reported metrics, surpassing the strongest baseline by about 2.9\% on average in Scenario\#1 and by about 3.4\% in Scenario\#2.
This demonstrates that explicitly constructing a stable alignment pathway for sparse satisfaction signals leads to more effective optimization toward user satisfaction.

(2) Secondly, introducing questionnaire-based satisfaction signals brings clear performance gains, and EASQ exploits them more effectively than existing SOTA methods.
Experimental results show that after incorporating questionnaire-based supervision, EMER$_\mathbf{S}$ achieves consistent improvements across major evaluation metrics.
More importantly, when compared with SAQRec, which also models questionnaire-based satisfaction signals, EASQ still achieves superior performance by 3.2\% on average. This indicates that our method not only leverages questionnaire signals, but also preserves their influence during training, preventing them from being overwhelmed by dominant behavioral signals.

(3) Finally, EASQ demonstrates strong generalization ability across multiple scenarios.
We conduct offline evaluations in multiple application scenarios, and EASQ consistently outperforms competing methods in all settings, achieving stable and significant improvements.
These results suggest that EASQ is not tailored to a specific scenario, but instead provides a generally effective framework for satisfaction-aware recommendation under sparse supervision.

\begin{figure*}[t]
    \centering
    \begin{tabular}{cccc}
        \includegraphics[width=0.23\textwidth]{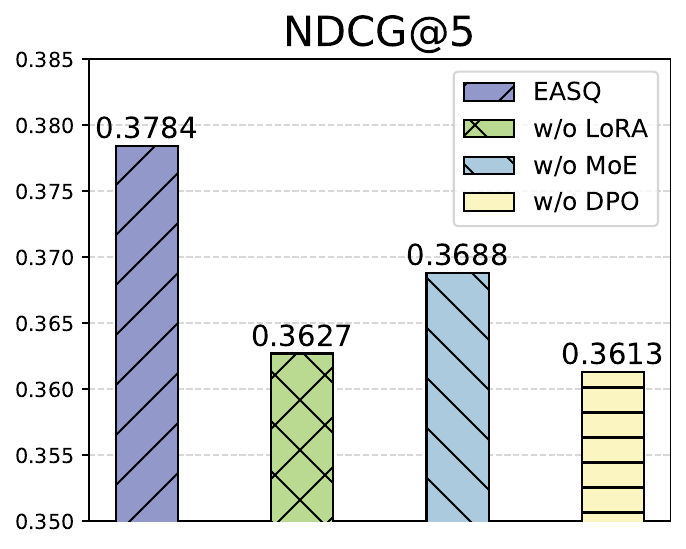} &
        \includegraphics[width=0.23\textwidth]{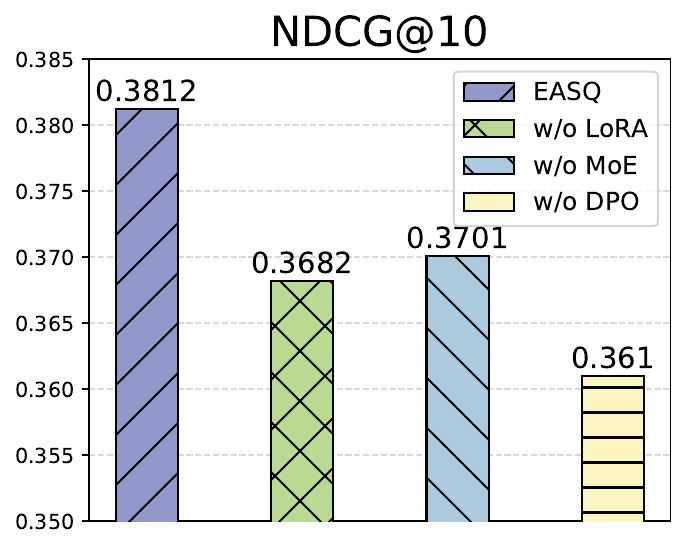} &
        \includegraphics[width=0.23\textwidth]{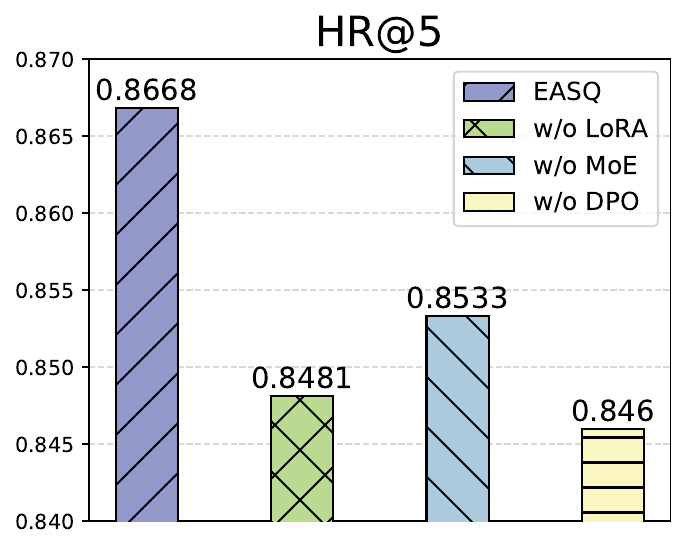} &
        \includegraphics[width=0.23\textwidth]{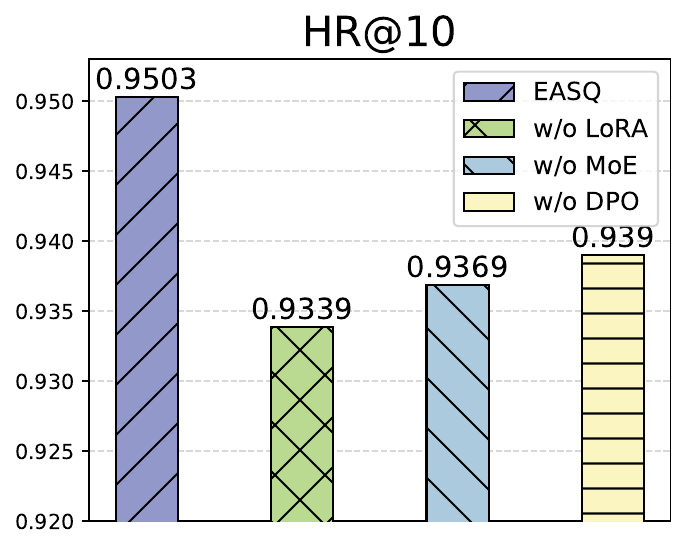} \\
    \end{tabular}
    \caption{Performance for Ablation Study on Scenario\#2.}
    \label{fig:abla}
\end{figure*}

\begin{figure*}[t]
    \centering
    \begin{tabular}{cccc}
        \subcaptionbox{$K_1$ on NDCG@5\label{fig:img1}}{\includegraphics[width=0.23\textwidth]{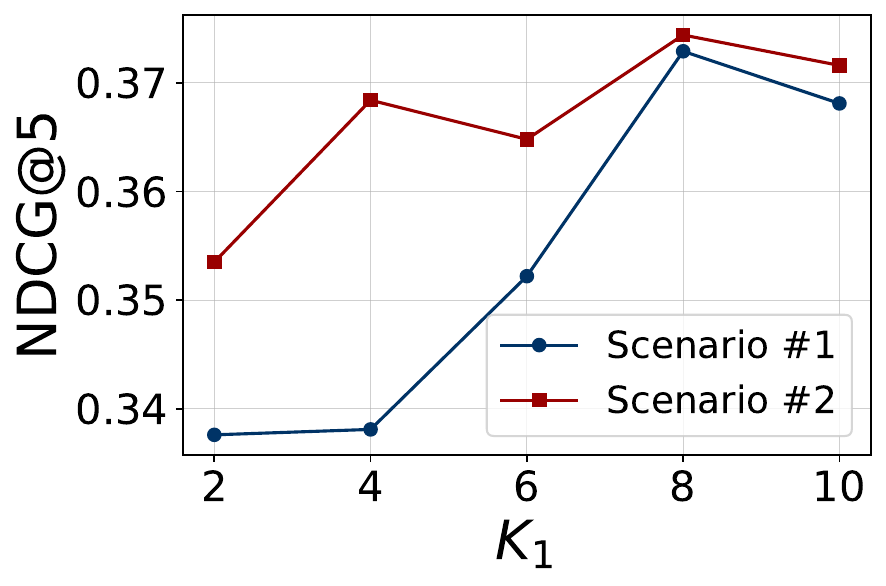}} &
        \subcaptionbox{$K_2$ on NDCG@5\label{fig:img4}}{\includegraphics[width=0.23\textwidth]{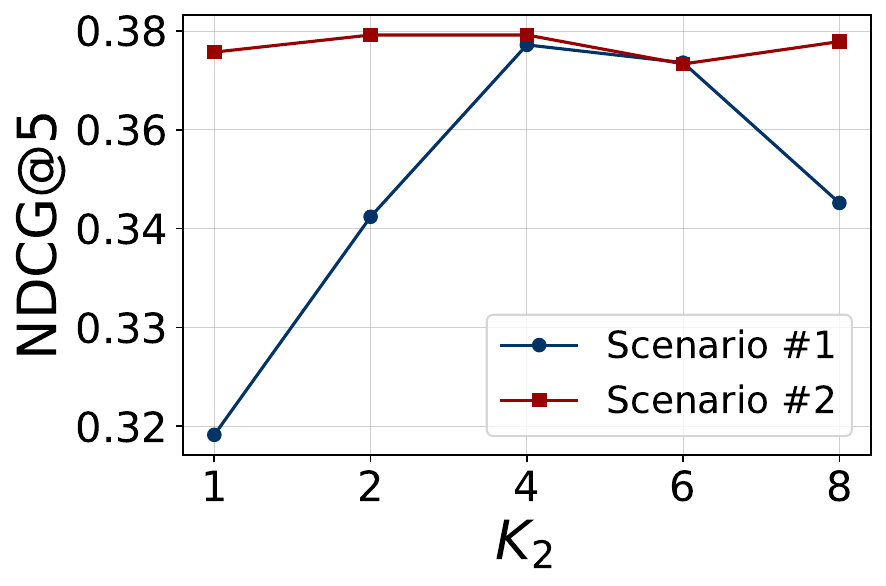}} &
        \subcaptionbox{$\lambda_1$ on NDCG@5\label{fig:img7}}{\includegraphics[width=0.23\textwidth]{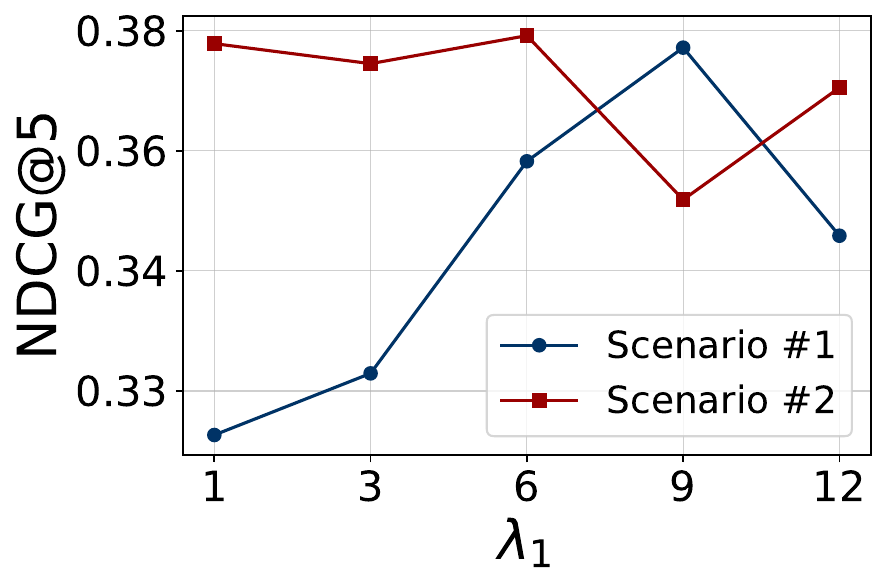}} &
        \subcaptionbox{$\lambda_2$ on NDCG@5\label{fig:img7}}{\includegraphics[width=0.23\textwidth]{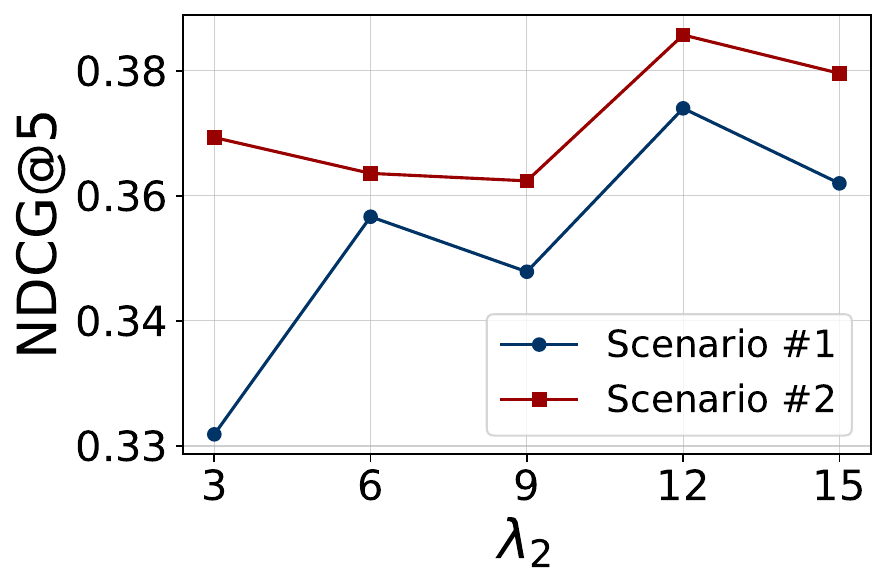}} \\
        \subcaptionbox{$K_1$ on HR@5\label{fig:img2}}{\includegraphics[width=0.23\textwidth]{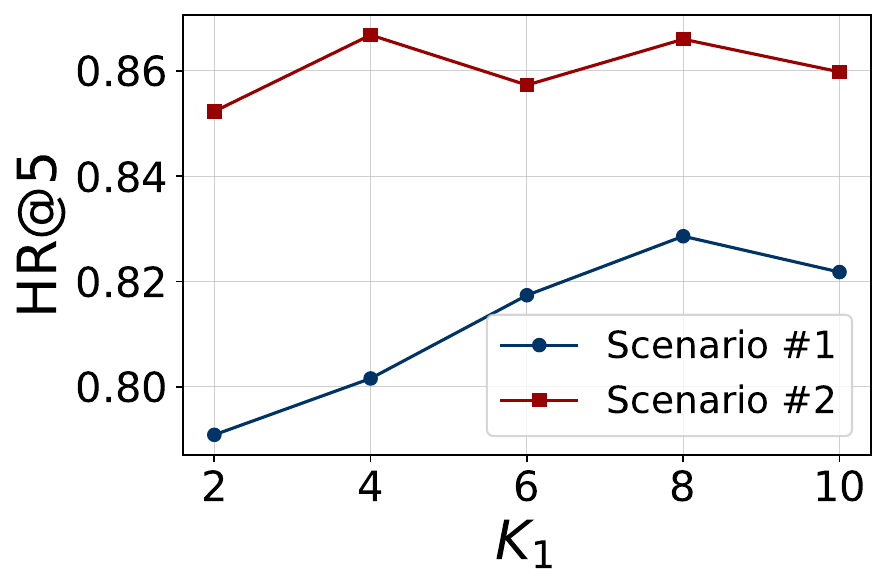}} &
        \subcaptionbox{$K_2$ on HR@5\label{fig:img5}}{\includegraphics[width=0.23\textwidth]{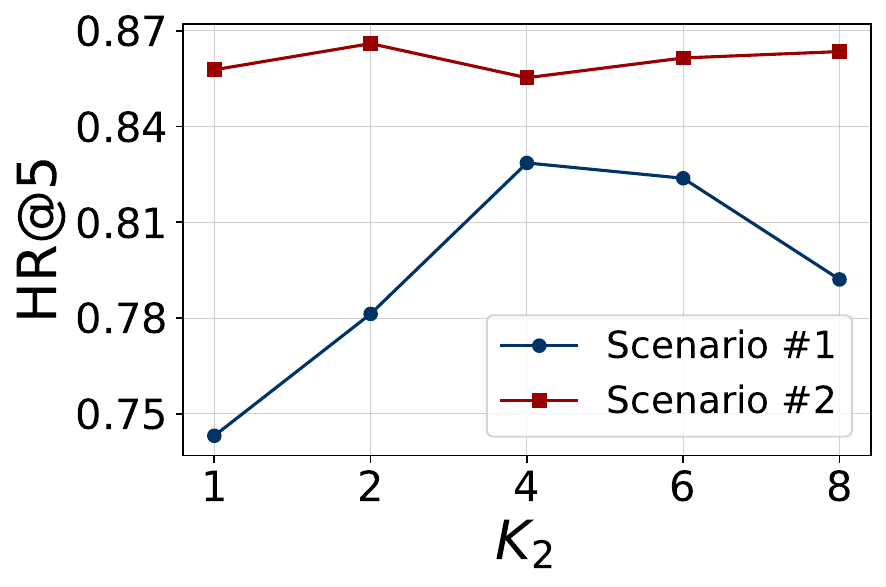}} &
        \subcaptionbox{$\lambda_1$ on HR@5\label{fig:img8}}{\includegraphics[width=0.23\textwidth]{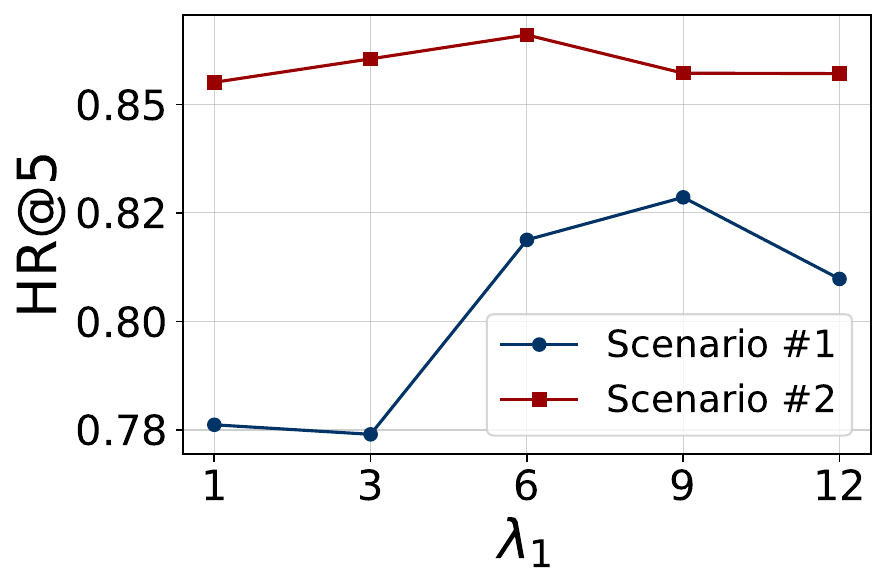}} &
        \subcaptionbox{$\lambda_2$ on HR@5\label{fig:img8}}{\includegraphics[width=0.23\textwidth]{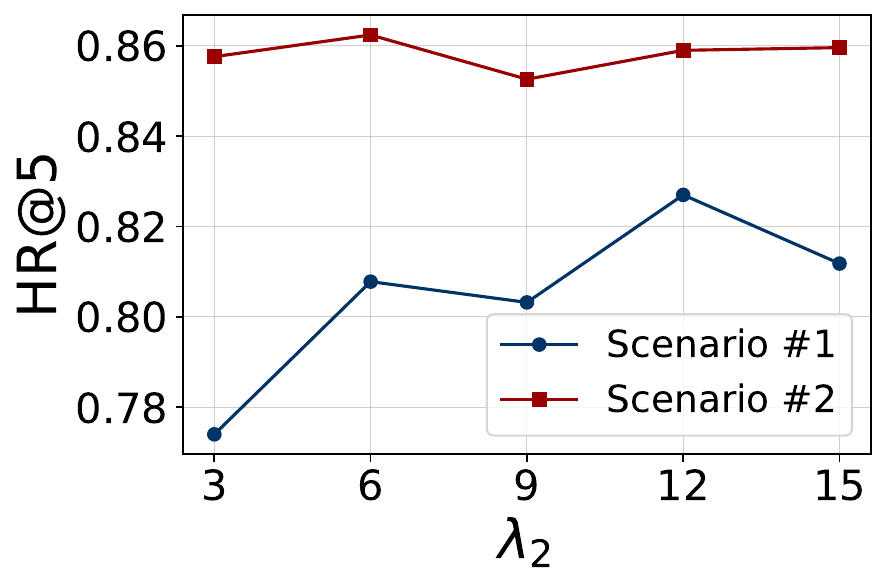}} \\
    \end{tabular}
    \caption{Sensitivity analysis for four key parameters, evaluated on NDCG@5 and HR@5 across two scenarios.
    % (a)(d) When studying $K_1$, we fix $K_2=2$ and $\lambda=6$; (b)(e) when studying $K_2$, we fix $K_1=8$ and $\lambda=6$; (c)(f) when studying $\lambda$, we fix $K_1=8$ and $K_2=2$.
    }
    \label{fig:3x3grid}
\end{figure*}

\subsection{Online A/B Testing}
To comprehensively evaluate EASQ in real-world recommendation scenarios, we conducted online A/B testing for 7 consecutive days across two scenarios on a short-video platform. 
We randomly allocated 5.1\% of the main traffic, with the control group using the online base model and the experimental group using the base model integrated with EASQ.

The evaluation covers a comprehensive set of user satisfaction metrics, including:
\begin{itemize}
    \item retention metric: 7-day Lifetime (LT7).
    \item implicit feedback metrics: APP stay time, watch time and video view.
    \item explicit feedback metrics: actions such as like, follow, comment, forward, etc.
    \item questionnaire-related metrics: positive and negative feedback ratio of questionnaires (Q-rate), questionnaire satisfaction (Q-Satisfied), and questionnaire dissatisfaction (Q-Dissatisfied).
\end{itemize}

As shown in Table~\ref{tab:online results}, EASQ consistently improves positive indicators while reducing negative ones compared to the baseline online model.

\textbf{Overall online performance gains.}
Compared with the strong production-level baseline, EASQ brings consistent and statistically significant improvements across core online metrics.
In particular, we observe clear gains on key engagement indicators such as APP stay time and watch time, indicating that aligning the model with sparse satisfaction signals leads to sustained user value rather than short-term behavioral optimization.

\textbf{Comprehensive improvements across multiple metric dimensions.}
Beyond retention and engagement, the proposed method achieves simultaneous improvements on both implicit and explicit feedback metrics in diverse recommendation scenarios.
EASQ not only increases video views and interaction-related behaviors (e.g., like, follow, and comment), but also effectively improves questionnaire-based satisfaction metrics.
Specifically, Q-rate and Q-Satisfied are consistently improved, while negative feedback indicators such as Q-Dissatisfied are reduced.
This demonstrates that EASQ enhances user experience in a holistic manner, rather than overfitting to a single class of online objectives.

\textbf{Robust effectiveness across multiple real-world scenarios.}
The observed performance gains are consistent across multiple short-video recommendation scenarios with different traffic distributions and user behaviors.
This suggests that EASQ provides a stable and transferable alignment mechanism, enabling effective adaptation of the base model to diverse online environments.
The robustness across scenarios further confirms the practicality of our approach for large-scale real-world deployment.

\subsection{Ablation Study}
To better understand the contribution of each component in EASQ, we conduct ablation studies by selectively removing key modules.
Specifically, we consider the following variants: 
\textbf{w/o LoRA}, which removes the low-rank adaptation module and feeds the frozen base model parameters into the satisfaction alignment task; 
\textbf{w/o MoE}, which replaces the mixture-of-experts architecture with a single expert network; 
and \textbf{w/o DPO}, which removes $\mathcal{L}_{\text{DPO}}$. At this time, the main task can only perceive satisfaction alignment information through the satisfaction signal injected by the low-level LoRA.

As shown in Figure~\ref{fig:abla}, removing any component leads to consistent performance degradation compared with the full EASQ model, indicating that all modules contribute positively to the final performance.
In particular, the w/o LoRA variant exhibits the most significant drop across both behavioral and satisfaction-related metrics, highlighting the importance of parameter-efficient adaptation under sparse satisfaction supervision.
Without LoRA, the model lacks sufficient flexibility to effectively align the base representations with satisfaction signals.

Replacing the MoE structure with a single expert also leads to clear performance degradation, demonstrating that expert-level specialization is crucial for modeling heterogeneous satisfaction patterns across different users and scenarios.
Without multiple experts, the model lacks sufficient capacity to disentangle diverse and potentially conflicting satisfaction signals, resulting in weaker representation and reduced alignment quality.

For the w/o DPO variant, removing the DPO-based alignment loss further degrades performance. 
This shows that directly aligning model outputs with questionnaire feedback through $\mathcal{L}_{\text{DPO}}$ is critical for effective satisfaction learning. 
When DPO is removed, the main ranking task can only receive satisfaction information indirectly through the low-level LoRA injection, which is insufficient to provide strong and stable alignment supervision, leading to weaker transfer of satisfaction signals to the final ranking results.

\subsection{Hyperparameter Analysis}
We investigate the impact of four key hyperparameters in EASQ: 
(1) the number of experts in main task $K_1$, 
(2) the number of experts in satisfaction alignment task $K_2$, 
(3) the weight of the satisfaction task loss $\lambda_1$,
(4) the weight of the alignment DPO loss $\lambda_2$,
All experiments are conducted by varying one hyperparameter at a time while keeping others fixed.

\subsubsection{Effect of the number of main-task experts $K_1$.}
As $K_1$ increases, all metrics initially show a clear improvement, indicating that a larger number of main-task experts helps the model capture more diverse user preferences.
However, beyond a certain point, the performance plateaus or slightly declines, likely due to over-parameterization, which increases model complexity without providing additional useful information and may lead to overfitting.

\subsubsection{Effect of the number of satisfaction alignment task experts $K_2$.}
The trends for $K_2$ are similar to those observed for $K_1$, with metrics improving as the number of satisfaction alignment task experts increases.
However, the optimal $K_2$ is smaller than that of $K_1$, likely because the questionnaire-based satisfaction signals are relatively sparse, and using too many experts may lead to overfitting or insufficient supervision for each expert.

\subsubsection{Effect of the satisfaction task loss weight $\lambda_1$.}
The impact of the alignment loss weight $\lambda_1$ varies across scenarios.
Each scenario exhibits a different optimal $\lambda_1$, indicating that the relative importance of aligning the main-task and satisfaction-task outputs depends on the characteristics of the underlying data.

\subsubsection{Effect of the alignment DPO loss weight $\lambda_2$.}
As $\lambda_2$ increases, the overall performance consistently improves, indicating that stronger alignment with satisfaction signals benefits the ranking model. 
However, when $\lambda_2$ exceeds a certain threshold, performance starts to degrade. 
This suggests that overly emphasizing the alignment objective may weaken the contribution of the main behavioral learning task, leading to suboptimal ranking quality. 
Therefore, an appropriate balance between the main task loss and the alignment loss is crucial for achieving optimal performance.

\section{Conclusion}
In this work, we propose EASQ, a novel framework that enables real-time alignment of ranking models with true user satisfaction by integrating sparse questionnaire feedback into online learning. Through an independent parameter pathway built with a multi-task architecture and a lightweight LoRA module, EASQ effectively prevents sparse satisfaction supervision from being overwhelmed by dense behavioral signals. The parameter-isolated LoRA injection allows targeted adaptation without destabilizing the backbone, while the DPO-based online optimization objective directly aligns model outputs with satisfaction signals in real time. Extensive offline evaluations and large-scale online A/B tests demonstrate that EASQ consistently improves satisfaction-related metrics across multiple scenarios. Its successful deployment in a production short-video recommendation system further validates its practicality and effectiveness in delivering stable and significant business gains.

\bibliographystyle{ACM-Reference-Format}
\bibliography{reference}

% %%
% %% If your work has an appendix, this is the place to put it.
% \appendix









\end{document}